\documentclass[useAMS,usenatbib,usegraphicx]{mn2e}
\begin{document}

  \title[Pulsar Timing Array]{On measuring the gravitational-wave background using Pulsar Timing Arrays}
  
  \author[van Haasteren et al.]{Rutger van Haasteren$^1$, Yuri Levin$^{1,2}$,
    Patrick McDonald$^3$, Tingting Lu$^{3,4}$
  \\
    $^1$Leiden University, Leiden Observatory, P.O. Box 9513, NL-2300 RA
      Leiden, the Netherlands
  \\
    $^2$Leiden University, Lorentz Institute, P.O. Box 9506, NL-2300 RA Leiden,
      the Netherlands
  \\
    $^3$CITA, 60 St.~George Street, Toronto, Ontario M5S 3H8, Canada
  \\
    $^4$Department of Astronomy, University of Toronto, 60 St.~George Street,
      Toronto, Ontario M5S 3H8, Canada}
 
  \date{printed \today}

  \maketitle

%-- ABSTRACT -------------------------------------------------------------------
  \begin{abstract}
    Long-term precise timing of Galactic millisecond pulsars holds great promise
    for measuring the long-period (months-to-years) astrophysical gravitational
    waves. Several gravitational-wave observational programs, called Pulsar
    Timing Arrays (PTA), are being pursued around the world.
    
    Here we develop a Bayesian algorithm for measuring the stochastic
    gravitational-wave background (GWB) from the PTA data. Our algorithm has
    several strengths: (1) It analyses the data without any loss of information,
    (2) It trivially removes systematic errors of known functional form,
    including quadratic pulsar spin-down, annual modulations and jumps due to a
    change of equipment, (3) It measures simultaneously both the amplitude and
    the slope of the GWB spectrum, (4) It can deal with unevenly sampled data
    and coloured pulsar noise spectra. We sample the likelihood function using
    Markov Chain Monte Carlo (MCMC) simulations. We extensively test our
    approach on mock PTA datasets, and find that the algorithm has significant
    benefits over currently proposed counterparts. We show the importance of
    characterising all red noise components in pulsar timing noise by
    demonstrating that the presence of a red component would significantly
    hinder a detection of the  GWB

    Lastly, we explore the dependence of the signal-to-noise ratio on the
    duration of the experiment, number of monitored pulsars, and the magnitude
    of the pulsar timing noise. These parameter studies will help formulate
    observing strategies for the PTA experiments. 
\end{abstract}
%-- abstract -------------------------------------------------------------------


%-- MAIN TEXT ------------------------------------------------------------------
  \section{Introduction}
    At the time of this writing several large projects are being pursued in
    order to directly detect astrophysical gravitational waves. This paper
    concerns a program to detect gravitational waves using pulsars as
    nearly-perfect Einstein clocks.  The practical idea is to time a set of
    millisecond pulsars (called the ``Pulsar Timing Array'', or PTA) over a
    number of years \citep{Foster}. Some of the millisecond pulsars create pulse
    trains of exceptional regularity. By perturbing the space-time between a
    pulsar and the Earth, the gravitational waves (GWs) will cause extra
    deviations from the periodicity in the pulse arrival times \citep{Estabrook,
    Sazhin, Detweiler}. Thus from the measurements of these deviations (called
    ``timing-residuals'', or TR), one may measure the gravitational waves.
    Currently, several PTA project are operating around the globe. Firstly, at
    the Arecibo Radio Telescope in North-America several millisecond pulsars
    have been timed for a number of years. These observations have already been
    used to place interesting upper limits on the intensity of gravitational
    waves which are passing through the Galaxy \citep{Kaspi, Lommen}. Together
    with the Green Bank Telescope, the Arecibo Radio Telescope will be used as
    an instrument of NANOGrav, the North American PTA. Secondly, the European
    PTA is being set up as an international collaboration between Great Britain,
    France, Netherlands, Germany, and Italy, and will use 5 European radio
    telescopes to monitor about 20 millisecond pulsars \citep{Stappers}.
    Finally, the Parkes PTA in Australia has been using the Parkes multi-beam
    radio-telescope to monitor 20 millisecond pulsars \citep{Manchester}. Some
    of the Parkes and Arecibo data have also been used to place the most
    stringent limits on the GWB to date \citep{Jenet-2006}.

    One of the main astrophysical targets of the PTAs is the stochastic
    background of the gravitational waves (GWB). This GWB is thought to be
    generated by a large number of black-hole binaries which are thought to be
    located at the centres of galaxies \citep{Begelman, Phinney, Jaffe, Wyithe,
    Sesana}, by relic gravitational waves \citep{Grishchuk}, or, more
    speculatively, by cusps in the cosmic-string loops \citep{Damour}. This
    paper develops an algorithm for the optimal PTA measurement of such a GWB.
 
    The main difficulty of such a measurement is that not only Gravitational
    Waves  create the pulsar timing-residuals. Irregularities of the pulsar-beam
    rotation (called the ``timing noise''), the receiver noise, the imprecision
    of local clocks, the polarisation calibration of the telescope
    \citep{Britton}, and the variation in the refractive index of the
    interstellar medium all contribute significantly to the timing-residuals,
    making it a challenge to separate these noise sources from the
    gravitational-wave signal. However, the GWB is expected to induce
    correlations between the timing-residuals of different pulsars. These
    correlations are of a specific functional form [given by
    Eq.~(\ref{eq:alphaab}) below], which is different from those introduced by
    other noise sources \citep{Hellings}.  \citet[hereafter J05]{Jenet-2005}
    have invented a clever algorithm which uses the uniqueness of the
    GWB-induced correlations to separate the GWB from other noise sources, and
    thus to measure the magnitude of the GWB.  Their idea was to measure the
    timing-residual correlations for all pairs of the PTA pulsars, and check how
    these correlations depend on the sky-angles between the pulsar pairs. J05
    have derived a statistic which is sensitive to the functional form of the
    GWB-induced correlation; by measuring the value of this statistic one can
    infer the strength of the GWB.  While J05 algorithm appears robust, we
    believe that in its current form it does have some drawbacks, in particular:
    \newline
    (1) The statistic used by J05 is non-linear and non-quadratic in the
    pulsar-timing-residuals, which makes its statistical properties
    non-transparent.\newline
    (2) The pulsar pairs with the high and low intrinsic timing noise make
    equal contributions to the J05 statistic, which is clearly not
    optimal.\newline
    (3) The J05 statistic assumes that the timing-residuals of all the PTA
    pulsars are measured during each observing run, which is generally not the
    case.\newline
    (4) The J05 signal-to-noise analysis relies on the prior knowledge of the
    intrinsic timing noise, and there is no clean way to separate this timing
    noise from the GWB.\newline
    (5) The prior spectral information on GWB is used for whitening the signal;
    however, there is no proof that this is an optimal procedure. The spectral
    slope of the GWB is not measured.

    In this paper we develop an algorithm which addresses all of the problems
    outlined above.  Our method is based on essentially the same idea as that of
    J05: we use the unique character of the GWB-induced correlations to measure
    the intensity of the GWB. The algorithm we develop below is Bayesian, and by
    construction uses optimally all of the available information. Moreover, it
    deals correctly and efficiently with all systematic contributions to the
    timing-residuals which have a known functional form, i.e.~the quadratic
    pulsar spin-downs, annual variations, one-time discontinuities (jumps) due
    to equipment change, etc. Many parameters of the timing model (the model
    popular pulsar timing packages use to generate TRs from pulsar arrival
    times) fall in this category.

    The plan of the paper is as follows. In the next section we review the
    theory of the GWB-generated timing residuals and introduce our  model for
    other contributions to the timing residuals.  In Sec.~\ref{sec:bayes} we
    explain the principle of Bayesian analysis for GWB-measurement with a  PTA,
    and we evaluate the Bayesian likelihood function. There we also show how to
    analytically marginalise over the contributions of known functional form but
    unknown amplitude (i.e., annual variations, quadratic residuals due to
    pulsar spin-down, etc.). The details of this calculation are laid out in
    Appendix A. Section~\ref{sec:montecarlo} discusses the numerical integration
    technique which we use in our likelihood analysis: the Markov Chain Monte
    Carlo (MCMC). In Sec.~\ref{sec:results} we show the analyses of mock PTA
    datasets.  For each mock dataset, we construct the probability distribution
    for the intensity of the GWB, and demonstrate its consistency with the input
    mock data parameters. We study the sensitivity of our algorithm for
    different PTA configurations, and investigate the dependence of the
    signal-to-noise ratio on the duration of the experiment, on redness and
    magnitude of the pulsar timing noise, and on the number of clocked pulsars.
    In Sec.~\ref{sec:conclusions} we summarise our results.

%-------------------------------------------------------------------------------

  \section{The Theory of GW-generated timing-residuals} \label{sec:theory}
    \subsection{Timing residual correlation} \label{sec:correlation}
      The measured millisecond-pulsar timing-residuals  contain
      contributions from several stochastic and deterministic processes. The
      latter include the gradual deceleration of the pulsar spin, resulting in a
      pulsar rotational period derivative which induces timing residuals varying
      quadratically with time (hereafter referred to as ``quadratic spin-down''),
      the annual variations due to the imperfect knowledge of the pulsar
      positions on the sky, the ephemeris variations caused by the known planets
      in the solar system, and the jumps due to equipment change (Manchester
      2006). The stochastic component of the timing-residuals will be caused by
      the receiver noise, clock noise, intrinsic timing noise, the refractive
      index fluctuations in the interstellar medium, and, most importantly for
      us, the GWB. For the purposes of this paper we restrict ourselves to
      considering the quadratic spin-downs, intrinsic timing noise, and the GWB;
      other components can be similarly included, but we omit them for
      mathematical simplicity. In this case, the $i^{\hbox{\tiny{th}}}$ timing
      residual of the $a^{\hbox{\tiny{th}}}$ pulsar can be written as  
      \begin{equation}
	\delta t_{ai}=\delta t_{ai}^{\rm GW}+\delta t_{ai}^{\rm PN}+Q(t_{ai}),
	\label{eqtai}
      \end{equation}
      where $\delta t_{ai}^{\rm GW}$  and $\delta t_{ai}^{\rm PN}$ are caused
      by the GWB and the pulsar timing noise, respectively, and 
      \begin{equation}
	Q_a(t_{ai})=A_{a1}+A_{a2} t_{ai}+A_{a3} t_{ai}^2
	\label{Qa}
      \end{equation}
      represent the quadratic spin-down. One expects the timing noise from
      different pulsars to be uncorrelated, while the GWB will cause
      correlations in the timing-residuals between different pulsars. Therefore,
      the information about GWB can be extracted by correlating the timing
      residual data between the different pulsars (J05). If one assumes that
      both GWB-generated residuals and the intrinsic timing noise are stochastic
      Gaussian processes, then we can represent them by the $(n \times n)$
      coherence matrices:
      \begin{eqnarray} 
	\label{coherence}
	\langle\delta t_{ai}^{\rm GW}\delta t_{bj}^{\rm GW}\rangle&=&C^{\rm GW}_{(ai)(bj)}
	  \nonumber\\
	\langle\delta t_{ai}^{\rm PN}\delta t_{bj}^{\rm PN}\rangle&=&C^{\rm PN}_{(ai)(bj)},
      \end{eqnarray}
      with the total coherence matrix given by
      \begin{equation}
      C_{(ai)(bj)}=C^{\rm GW}_{(ai)(bj)}+C^{PN}_{(ai)(bj)}.
      \label{cfull}
      \end{equation} 
      The timing-residuals are then distributed as a multidimensional Gaussian:
      \begin{eqnarray} \label{eq:gaussian}
	P\left(\vec{\delta t}\right)&=&\frac{1}{\sqrt{\left(2\pi\right)^{n}
	  \det C}} \exp\left[-\frac{1}{2}\sum_{{(ai)(bj)}}
	  (\vec{\delta t}_{(ai)}-Q_a(t_{ai}))\right.\nonumber\\
	& &\left.C^{-1}_{(ai)(bj)} (\vec{\delta t}_{(bj)}-Q_b(t_{bj}))\right],
      \end{eqnarray}
      where $P$ denotes the probability distribution of the timing-residuals.
      To be able to use Eq.~(\ref{eq:gaussian}) we must\newline
      (1) be able to evaluate the GWB-induced coherence matrix from
      the theory, as a function of variables that parametrise the GWB spectrum,
      and \newline
      (2) introduce well-motivated parametrization of the pulsar timing noise. 
      In this work, the spectral density of the stochastic GW
      background is taken to be a power law \citep{Phinney, Jaffe, Wyithe, Maggiore}
      \begin{equation} \label{eq:spectraldensity}
	S_h=A^2\left(f\over \hbox{yr}^{-1}\right)^{-\gamma},
      \end{equation}
      where $S_h$ represents the spectral density, $A$ is the GW amplitude, $f$
      is the GW frequency, and $\gamma$ is an exponent characterising the GWB
      spectrum. If the GWB is dominated by the supermassive black hole binaries,
      then $\gamma=7/3$ (Phinney 2001). This definition is equivalent to the use
      of the characteristic strain as defined in \citet{Jenet-2006}:
      \begin{equation} \label{eq:characteristicstrain}
	h_c=A\left(f\over \hbox{yr}^{-1}\right)^{\alpha},
      \end{equation}
      with $\gamma = 1 - 2\alpha$. The GWB-induced coherence matrix is
      then given by
      \begin{eqnarray}
	C^{\rm GW}_{(ai)(bj)} &=&
	  \frac{A^2\alpha_{ab}}{\left(2\pi\right)^{2}f_{L}^{1+\gamma}}
	  \left\{ \Gamma(-1-\gamma)\sin\left(\frac{-\pi\gamma}{2}\right)
	  \left(f_{L}\tau\right)^{\gamma+1} \right.\nonumber\\
	& &-\left.\sum_{n=0}^{\infty}\left(-1\right)^{n}
	  \frac{\left(f_{L}\tau\right)^{2n}}{(2n)!
	  \left(2n-1-\gamma\right)}\right\} . 
	\label{eq:CGW}
      \end{eqnarray}
      Here $\alpha_{ab}$ is the geometric factor given by
      \begin{equation}
	\alpha_{ab}={3 \over 2}\frac{1-\cos\theta_{ab}}{2}
	  \ln\left(\frac{1-\cos\theta_{ab}}{2}\right)
	  -\frac{1}{4}\frac{1-\cos\theta_{ab}}{2}+\frac{1}{2}
	  +\frac{1}{2}\delta_{ab},
	\label{eq:alphaab}
      \end{equation}
      where  $\theta_{ab}$ is the angle between pulsar $a$ and pulsar $b$
      \citep{Hellings}, $\tau=2\pi\left(t_{ai}-t_{bj}\right)$,
      $\Gamma$ is the gamma function, and $f_L$ is the low cut-off frequency,
      chosen so that $1/f_L$ is much greater than the duration of the PTA
      operation.  Introducing $f_L$ is a mathematical necessity, since
      otherwise the GWB-induced correlation function would diverge.  However,
      we show below that the low-frequency part of the GWB is indistinguishable
      from an extra  spin-down of all pulsars which we already correct for, and
      that our results do not depend on the choice of $f_L$ provided that
      $f_L\tau\ll 1$.

      The pulsar timing noise is assumed to be Gaussian, with a certain
      functional form of the power spectrum.  The true profile of the
      millisecond pulsar timing noise spectrum is not well-known at present
      time. The timing residuals of the most precisely observed pulsars indicate
      that pulsar timing noise has a white and poorly-constrained red component
      (J.~Verbiest and G.~Hobbs, private communications).  

      For the purposes of this paper we will always choose the spectra to be
      of the same functional form for all pulsars, but this is not an inherent limitation of the
      algorithm. We  consider 3 cases of pulsar timing noise
      spectra:\newline
      (1) White (flat) spectra\newline
      (2) Lorentzian spectra\newline
      (3) Power-law spectra\newline
      Obviously, one could also consider a timing noise which is a superposition of these components;
      we do not do this at this exploratory stage.
      If we choose the pulsar timing noise spectrum to be white, with an
      amplitude $N_a$, the resulting correlation matrix becomes:
      \begin{eqnarray}
	C^{\rm PN-white}_{(ai)(bj)} &=& N_a^2\delta_{ab}\delta_{ij}.
	\label{eq:CPN-white}
      \end{eqnarray}

      The Lorentzian spectrum is a red spectrum with a typical frequency that
      determines the redness of the timing noise:
      \begin{equation}
	S_{a}(f)=\frac{N_{a}^{2}}{f_0\left(1+\left(f\over f_0\right)^2\right)},
	\label{eq:SPN-lorentzian}
      \end{equation}
      which yields the following correlation matrix:
      \begin{eqnarray}
	C^{\rm PN-lor}_{(ai)(bj)} &=&
	  N_a^2\delta_{ab}\exp\left(-f_0\tau\right),
	\label{eq:CPN-lorentzian}
      \end{eqnarray}
      where $f_0$ is a typical frequency and $N_a$ is the amplitude.

      By using a power law spectral density with amplitude $N_a$ and
      spectral index $\gamma_a$, one  gets a timing-noise coherence matrix analogous 
      to the one in Eq.~(\ref{eq:CGW}):
      \begin{eqnarray}
	C^{\rm PN-pl}_{(ai)(bj)} &=& \frac{N_a^2\delta_{ab}}
	  {f_{L}^{\gamma_a-1}}\left\{
	  \Gamma(1-\gamma_a)\sin\left(\frac{\pi\gamma_a}{2}\right) 
	  \left(f_{L}\tau\right)^{\gamma_a-1} \right.\nonumber\\
	& &-\left.\sum_{n=0}^{\infty}\left(-1\right)^{n}
	  \frac{\left(f_{L}\tau\right)^{2n}}
	  {(2n)!\left(2n+1-\gamma_a\right)}\right\} . 
	\label{eq:CPN-power}
      \end{eqnarray}

  \section{Bayesian approach} \label{sec:bayes}
    \subsection{Basic ideas}
      The method described in this report is based upon a Bayesian approach to
      the parameter inference.  The general idea of the method is to (a) assume
      that the physical processes which produce the timing-residuals can be
      characterised by several parameters, and (b) use the Bayes theorem to
      derive from the measured data the probability distribution of the
      parameters of our interest. In our case, we assume that the timing
      residuals are created by \newline
      (1) the GWB; we parametrise it by its amplitude $A$ and slope $\gamma$, as
      in equation (\ref{eq:spectraldensity}).\newline
      (2) the intrinsic timing noise of the 20 monitored millisecond pulsars. We
      assume that the timing noise of each of the pulsars is the random Gaussian
      noise, with a variety of possible spectra described in the previous
      section. We shall refer to the variables parametrizing the timing noise
      spectral shape as $TN_a$.\newline
      (3). The quadratic spin-downs, parametrised for each of the pulsars by
      $A_{a1}$, $A_{a2}$, and $A_{a3}$, cf.~Eq.~(\ref{Qa}).\newline

      With these assumptions, we shall  write down below the expression for the
      probability distribution $P({\rm data}|{\rm parameters})$ of the data, as
      a function of the parameters. By Bayes theorem, we can then compute the
      posterior distribution function; the probability distribution of the
      parameters given a certain dataset:

      \begin{eqnarray}
	\label{bayes}
	P(\hbox{parameters}| {\rm data})&=&
	  P({\rm data}|\hbox{parameters})\times\\
	  & &\times {P_0(\hbox{parameters})\over P({\rm data})}.\nonumber
      \end{eqnarray}
      Here $P_0({\rm parameters})$ is the prior probability  of the unknown
      parameters, which represents all our current knowledge about these
      parameters, and $P({\rm data})$ is the Bayesian evidence, which we will
      use here as a normalisation factor to ensure that $P(A,\gamma, TN_a,
      A_{a1}, A_{a2}, A_{a3}| {\rm data})$ integrates to unity over the
      parameter space. We note here that the Bayesian evidence is in essence a
      goodness of fit measure that can be used for model selection. However, we
      will ignore the Bayesian evidence in this work and postpone the model
      selection part of the algorithm to future work. For our purposes, we are
      only interested in $A$ and $\gamma$, which means that we have to integrate
      $P(A,\gamma, TN_a, A_{a1}, A_{a2}, A_{a3}| {\rm data})$ over all of the
      other parameters.  Luckily, as we show below, for a uniform prior the
      integration over $A_{a1}$, $A_{a2}$, and $A_{a3}$ can be performed
      analytically. This amounts to the {\it removal} of the quadratic spin-down
      component to the pulsar data. We emphasise that this removal technique is
      quite general, and can be readily applied to unwanted signal of any known
      functional form (i.e., annual modulations, jumps, etc.---see
      Sec.~\ref{sec:qsdremoval}), even if those parameters have already been fit
      for while calculating the timing residuals. The integration over $TN_a$
      must be performed numerically.
      
      In this work we shall use MCMC simulation as a multi dimensional
      integration technique. Besides flat priors for most of the parameters, we
      will use slightly peaked priors for parameters which have non-normalisable
      likelihood functions. This ensures that the Markov Chain can converge.

      In the rest of the paper, we detail the implementation and tests of our
      algorithm.

    \subsection{Removal of the quadratic spin-downs and other systematic
      signals of known functional form} \label{sec:qsdremoval}
      While this subsection  is written with the PTA in mind, it may well be
      useful for other applications in pulsar astronomy.  We thus begin with a
      fairly general discussion, and then make it more specific for the PTA
      case.
 
      Consider a random Gaussian process $\delta x^{\rm G}_i$ with a coherence
      matrix $C(\sigma)$, which is contaminated by several systematic signals
      with known functional forms $f_p(t_i)$ but a-priori unknown amplitudes
      $\xi_p$. Here $\sigma$ is a set of interesting parameters which we want to
      determine  from the data $\delta x$. The resulting signal is given by
      \begin{equation}
	\delta x_i=\delta x^{\rm G}_i+\sum_p \xi_p f_p(t_i),
	\label{deltareal1}
      \end{equation}
      or, in the vector form, by
      \begin{equation}
	\vec{\delta x}=\vec{\delta x}^{\rm G}+F\vec{\xi}.
	\label{deltareal2}
      \end{equation}
      Here the components of the vectors $\vec{\delta x}$, $\vec{\delta x}^{\rm
      G}$, and $\vec{\xi}$ are given by $\delta x_i$, $\delta x^{\rm G}_i$, and
      $\xi_p$, respectively, and $F$ is the non-square matrix with the elements
      $F_{ip}=f_p(t_i)$. Note that the dimensions of $\vec{\delta x}$ and
      $\vec{\xi}$ are different. The Bayesian probability distribution for the
      parameters is given by
      \begin{eqnarray}
	P(\sigma, \vec{\xi}|\vec{\delta
	x})&=& {M\over \sqrt{\det{C}}}\exp \left[-{1\over 2}
	(\vec{\delta x}-F\vec{\xi}) C^{-1} (\vec{\delta x}-F\vec{\xi}) \right]
	\nonumber\\
	& &\times P_0(\sigma, \vec{\xi}),
	\label{eq:pb1}
      \end{eqnarray}
      where $P_0$ is the prior probability and $M$ is the normalisation. Since
      we are only interested in $\sigma$, we can integrate $P(\sigma,
      \vec{\xi}|\vec{\delta x})$ over the variables $\vec{\xi}$. This process is
      referred to as  marginalisation; it can be done analytically if we assume
      a flat prior for $\vec{\xi}$ [i.e., if $P_0(\sigma,\vec{\xi})$ is
      $\vec{\xi}$-independent], since $\xi_p$ enter at most quadratically into
      the exponential above. After some straightforward mathematics which we
      have detailed in Appendix A, we get
      \begin{eqnarray}
	\label{premoved}
	P(\sigma|\vec{\delta x})&=&{M'\over \sqrt{\det(C)
	      \det(F^{\rm T} C^{-1} F)}}\\ & &\times \exp\left[-{1\over
	      2}\vec{\delta x}\cdot C^{\prime} \vec{\delta x}\right],\nonumber
      \end{eqnarray}
      where $M'$ is the normalisation, and
      \begin{equation}
	C^{\prime}=C^{-1}-C^{-1}F(F^{\rm T} C^{-1}F)^{-1}F^{\rm T}C^{-1},
	\label{C'}
      \end{equation}
      and the $T$-superscript stands for the transposed matrix. Equation
      (\ref{premoved}) is one of the main equations of the paper, since it
      provides a statistically rigorous way to remove (i.e., marginalise over)
      the unwanted systematic signals from random Gaussian processes.  One can
      check directly that the above expression for $ P(\sigma|\vec{\delta x})$
      is insensitive to the values $\xi_p$ of the amplitudes of the systematic
      signals in the Eq.~(\ref{deltareal1}).

      We now apply this formalism to account for  the quadratic spin-downs in
      the PTA. As in Sec.~\ref{sec:theory}, it will be convenient to use the
      2-index notation for the timing-residuals, $\delta t_{ai}$ measured at the
      time $t_{ai}$, where $a$ is the pulsar index, and $i$ is the number of the
      timing residual measurement for pulsar $a$. The space of the spin-down
      parameters $A_{aj}$, $j=1,2,3$ has $3N$ dimensions, where $N$ is the
      number of pulsars in the array. In the component language, we write
      \begin{equation}
	\delta t_{ai}=\delta t^{\rm G}_{ai}+\sum_{b, j}
	F_{(ai)(bj)} A_{bj},
	\label{Fnew}
      \end{equation}
      where
      \begin{equation}
	F_{(ai)(bj)}=\delta_{ab} t_{ai}^{j-1},
	\label{F1}
      \end{equation}
      $\delta t^G$ is the part of the timing residual due to a random Gaussian
      process (i.e., GWB, timing noise, etc.), and $j=1,2,3$. The quantities $
      F_{(ai)(bj)}$ are components of the matrix operator which acts on the
      $3N$-dimensional vector in the parameter space and produces a vector in
      the timing-residual space. For example, for 20 pulsars, each with 250
      timing residual observations, the matrix $F_{(ai)(bj)}$ has $20\times
      250=5000$ rows, each marked by 2 indices $a=1,..., 20$, $i=1,..., 250$,
      and $20\times 3=60$ columns, each marked by 2 indices $b=1,...,20$,
      $j=1,2,3$. Thus in the vector form, one can write Eq.~(\ref{Fnew}) as
      \begin{equation}
	\vec{\delta t}=\vec{\delta t}^{\rm G}+F\vec{A},
	\label{Fpta}
      \end{equation}
      which is identical to the Eq.~(\ref{deltareal2}). We thus can use
      Eq.~(\ref{premoved}) to remove the quadratic spin-down contribution from
      the PTA data.

      Although we only demonstrate this technique for quadratic spin-down, this
      removal technique will be useful for treating other noise sources in the
      PTA. All sources of which the functional form is known (and therefore can
      be fit for, as most popular pulsar timing packages do) can be dealt with,
      i.e.\newline
      (1) Annual variation of the timing-residuals due to the imprecise
      knowledge of the pulsar position on the sky.  The annual variation in each
      of the pulsars  will be a predictable function of the associated 2 small
      angular errors (latitude and longitude). Thus our parameter space will
      expand by 2N, but this will still keep the $F$ matrix manageable.\newline
      (2) Changes of equipment will introduce extra jumps, and must be taken
      into account. This is trivial to deal with using the techniques described
      above. \newline
      (3) Some of the millisecond pulsars are in binaries, and their orbital
      motion must be subtracted. The errors one makes in these subtractions will
      affect the timing-residuals. They can be parametrised and dealt with using
      the techniques of this section (we thank Jason Hessels for pointing this
      out).

    \subsection{Low-frequency cut-off}
      All predictions for GWB spectrum show a steep power law $\propto
      f^{-\gamma}$, where for black-hole binaries ${\gamma=7/3}$
      \citep{Phinney}. Physically, there is a low-frequency cut-off to the
      spectrum, due to the fact that black-hole binaries with periods greater
      than $~1000$ years shrink mostly due to the external friction (i.e.,
      scattering of circum-binary stars or excitation of density waves in a
      circum-binary gas disc), and not to gravitational radiation. However,
      while the exact value of the low-frequency cut-off is poorly constrained,
      the PTA should not be sensitive to it since the duration of the currently
      planned experiments is much shorter than $1000$ years. In this
      subsection, we show this formally by explicitly introducing the
      low-frequency cut-off and by demonstrating that our Bayesian
      probabilities are insensitive to its value. 

      Consider the expression in Eq.~(\ref{eq:CGW}) for the GWB-generated
      correlation matrix for the timing-residuals.  This expression contains an
      integral of the form
      \begin{equation}
	I=\int_{f_L}^\infty \cos(f\tau) f^{-(\gamma+2)} df,
	\label{I}
      \end{equation}
      where $\tau=2\pi (t_i-t_j)$. When the low-frequency cut-off is much
      smaller than the inverse of the experiment duration, i.e.~when
      $f_l\tau\ll 1$, the integral above can be expanded as
      \begin{equation}
	I=B\tau^{\gamma+1}+{1\over f_L^{\gamma+1}}\left\{ {1\over (\gamma+1)}
       	- {\left(f_L\tau\right)^2 \over 2(\gamma-1)} + \mathrm{O}\left[
       	(f_{L}\tau)^{4}\right]\right\},
      \label{I1}
      \end{equation}
      where 
      \begin{equation}
	B=\Gamma(-1-\gamma)\sin\left(\frac{-\pi\gamma}{2}\right)\tau^{\gamma+1}.
      \label{B}
      \end{equation}
      In the expansion above we have assumed $1<\gamma<3$. The terms which
      contain $f_L$ diverge when $f_L$ goes to zero, and scale as $\tau^0$ or
      $\tau^2$ with respect to the time interval. We now show that these
      divergent terms get absorbed in the process of elimination of the
      quadratic spin-downs.

      Suppose that we add to the timing-residuals of a pulsar a quadratic
      spin-down term, $A_1+A_2t+A_3 t^2$. The spin-down-removal procedure
      described in the previous section makes our results completely
      insensitive to this addition: $A$'s could be arbitrarily large but the
      measured GWB would still be the same. Clearly, the same is true if one
      treats $A_1$, $A_2$, $A_3$ not as fixed numbers, but as random variables
      drawn from some Gaussian distribution. The correlation introduced into
      the timing-residuals by adding a {\it random} quadratic spin-down is given
      by
      \begin{eqnarray}
      \langle\delta t_i \delta t_j\rangle&=&\langle A_1^2\rangle +\langle A_1
      A_2\rangle (t_i+t_j)\nonumber\\
			     &+&2\langle A_2^2\rangle t_it_j+\langle
			     A_1A_3\rangle (t_i^2+t_j^2)\label{nc}\\
			     &+&\langle A_2A_3\rangle
			     t_it_j(t_i+t_j)+\langle A_3^2\rangle t_i^2t_j^2.
			     \nonumber
      \end{eqnarray}
      The $f_L$-dependent part of Eq.~(\ref{I1}) contains terms which scale as
      $t_i^2+t_j^2$, $t_it_j$, and $const$, and thus have the same functional
      $t_i, t_j$ dependence as some of the terms in Eq.~(\ref{nc}). Since the
      terms in Eq.~(\ref{nc}) can be made arbitrarily large, it is clear that
      the terms corresponding to the low-frequency cutoff could be absorbed
      into the correlation function corresponding to the quadratic spin-down
      with the stochastic coefficients. We have made this argument for the
      timing-residuals from a single pulsar, but it is trivial to extend it to
      the case of multiple pulsars. Thus our results are not sensitive to the
      actual choice of the $f_L$ so long as $f_l\tau\ll 1$; this is confirmed
      by direct numerical tests.

  \section{Numerical integration techniques} \label{sec:montecarlo}
    \subsection{Metropolis Monte Carlo}
      The Bayesian probability distribution for the PTA is computed in
      multi dimensional parameter space, where all of the parameters except 2
      characterise intrinsic pulsar timing noise and other potential
      interferences. To obtain  meaningful information about the GWB, we need to
      integrate the probability function over all of the unwanted parameters.
      This is a challenging numerical task: a direct numerical integration over
      more than several parameters is prohibitively computationally expensive.
      Fortunately, numerical shortcuts do exist, and the most common among them
      is the Markov Chain Monte Carlo (MCMC) simulation. In a typical MCMC, a
      set of semi-random walkers sample the parameter space in a clever way,
      each generating a large number of sequential locations called a {\it
      chain} \citep{Newman}. After a sufficient  number of steps, the density of
      points of the chain becomes proportional to the Bayesian probability
      distribution. The number of steps required for the chain convergence
      scales linearly with the number of dimensions of the parameter space;
      typically few$\times 10^4$  steps are required for reliable convergence.
      In this paper we use the Metropolis \citep{Newman} algorithm for
      generating the chain, which can be used with an arbitrary distribution,
      the proposal distribution, for generating new locations of the chain.  We
      use a Gaussian proposal distribution, centred at the current location in
      the parameter space. During an initial period, the burn-in period, the
      width of the proposal distribution in all dimensional directions is set to
      yield the asymptotically optimal acceptance rate of $23.4\%$ for the
      Metropolis algorithm \citep{Roberts}. At the end of the MCMC simulation we
      check the convergence of the chain using the bootstrap method
      \citep{Efron}. We also calculate the global maximum likelihood value for
      all parameters using a conjugate directions search \citep{Brent}.

    \subsection{Current MCMC computational cost}
      The greatest computational challenge in constructing the chain is the
      fast evaluation of the matrix $C^{-1}$ in
      Eqs.~(\ref{premoved})\&(\ref{C'}).  If $250$ timing-residuals are
      measured for each of the pulsars (50 weeks for 5 years), the size of the
      matrix $C$ becomes $(5000\times 5000)$. We find it takes about $20$
      seconds to invert $C$ and thus about $1.5$ times as much to arrive to the
      next point in the chain. Therefore, for the required $~10^5$ chain points
      to get  the convergent distribution, we need of order $~1$ month of the
      single-processor computational time. On a cluster this can be done in a
      couple of days. We emphasize that this is an order $n^3$ process. For
      matrices of $(2000\times 2000)$ the calculation can be done overnight on
      a single modern workstation, but for $(10^4\times 10^4)$ the calculation
      is already a serious challenge.

      For the currently projected size of the datasets \citep{Manchester}, the
      amount of timing-residuals will most likely not exceed the $250$ (Hobbs,
      private communications). Thus, the brute-force method presented here is
      not computationally expensive for the projected data volume over the next
      5 years.

    \subsection{Choosing a suitable prior distribution} \label{sec:prior}
      For some models (e.g. the power law spectal density for pulsar timing
      noise) the likelihood function proves to be not normalisable. This would
      pose a serious problem in combination with uniform priors as the nuisance
      parameters then cannot be marginalised and the posterior cannot represent
      a probability distribution. Although this is a sign that our model is
      incorrect (infinite Bayesian Evidence/normalisation), this can be easily
      solved with a different parameterisation. We can always change coordinates
      in parameter space to a set for which all parameters have a finite domain,
      which guarantees that our likelihood function is normalisable. This
      procedure is equivalent to choosing a different prior (the Jacobian in the
      case of a coordinate transformation) for the original set. We therefore
      argue that we need to choose an appropriate prior for the non-normalisable
      parameters. We propose to use a Lorentzian shaped profile:
      \begin{equation}
	P_0(\gamma_i) = {\Delta_i \over \pi \left( \Delta_i^2 +
	\gamma_i^2 \right)},
	\label{eq:prior}
      \end{equation}
      where $\gamma_i$ is the parameter for which we are construction a prior,
      and $\Delta_i$ is some typical width/value for this parameter.

\begin{figure}
      \includegraphics[width=0.5\textwidth]{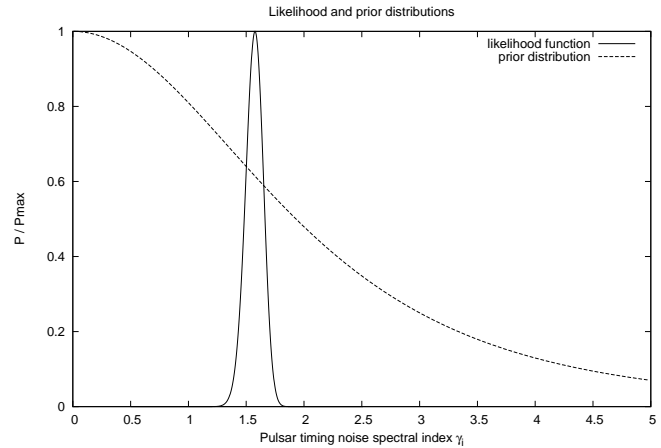}
      \caption{The likelihood and prior distribution for a pulsar timing noise
	spectral index parameters $\gamma_i$. The solid line represents the
	likelihood function. It is sharply peaked and it looks as if it drops to
	zero for high $\gamma_i$. However, for high $\gamma_i$ it will have a
	constant non-negligible value. The dashed line represents our chosen
	prior distribution. The prior is normalisable, and it's application
	makes the posterior distribution normalisable as well.}
	\label{fig:likelihood-prior}
      \end{figure}

      As an example we show the likelihood function and the prior for the pulsar
      timing noise spectral index parameter of Eq.~(\ref{eq:CPN-power}) in
      Fig.~\ref{fig:likelihood-prior}.  The likelihood function seems to drop to
      zero for high $\gamma_i$, but it actually has a non-negligible value for
      all $\gamma_i$ greater than the maximum likelihood value. The broadness of
      the prior is chosen such that it does not change the representation of the
      significant part of the likelihood in the posterior, but it does make sure
      that the posterior is normalisable.

  \subsection{Generating mock data} \label{sec:mockdata}
    In order to generate mock data, we produce a realization of  the
    multi dimensional Gaussian process of Eq.~(\ref{eq:gaussian}), as follows.
    We rewrite Eq.~(\ref{eq:gaussian}) is a basis in which $C$ is diagonal:
    \begin{equation}
      P\left( \vec{\delta t}\right)=\prod_{i=1}^{n}
      \frac{1}{\sqrt{\lambda_i}}\varphi\left( \frac{y_i}{\sqrt{\lambda_i}} \right),
      \end{equation}
    where,
    \begin{equation}
      \varphi(x) := \frac{1}{\sqrt{2\pi}}\exp\left(-\frac{x^2}{2}\right).
      \end{equation}
    Here $\lambda_i$ are the eigenvalues of $C$, and 
    \begin{equation}
      \vec{y}=T^{-1}\vec{\delta t}, 
      \label{diag1}
    \end{equation}
    where $T$ is the transformation matrix which diagonalizes $C$:
    \begin{equation}
      (T^{-1}CT)_{ij}=\lambda_i\delta_{ij}.
    \end{equation}
    Thus we follow the following steps:\newline (1) Diagonalize matrix $C$, find
    $T$ and $\lambda_i$.\newline (2) Choose $y_i$ from random gaussian
    distributions of widths $\sqrt{\lambda_i}$.\newline (3) Compute the timing
    residuals via Eq.~(\ref{diag1}).\newline

    It is then trivial to add deterministic processes, like quadratic
    spin-downs, to the simulated timing-residuals.

  \section{Tests and parameter studies} \label{sec:results}
    We test our algorithm by generating mock timing-residuals for a number of  millisecond
    pulsars which are positioned randomly in the sky. We found it convenient to
    parametrise the GWB spectrum by [cf.Eq.~(\ref{eq:spectraldensity})]
    \begin{equation}
      S_h(f)=A^2\left({f\over \hbox{yr}^{-1}}\right)^{-\gamma}.
      \label{eq:sd2}
    \end{equation}
    Our mock timing-residuals are a single realisation of GWB for some values of
    $A$ and $\gamma$ and the pulsar timing noise. Random quadratic-spin-down
    terms are added. We then perform several separate investigations as
    follows:\newline

    \subsection{Single dataset tests} \label{sec:singletests}
\begin{figure}
      \includegraphics[width=0.5\textwidth]{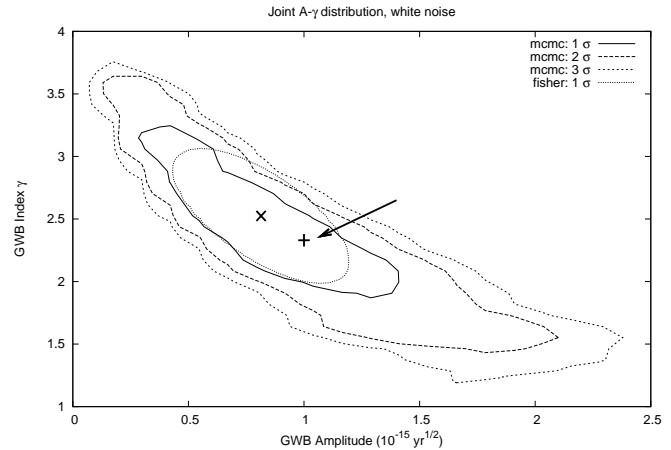}
      \caption{The GW likelihood function (GW amplitude, GW slope vs prob.
	  density contours), determined with the MCMC method for a set of mock
	  data with $20$ pulsars, and $100$ data points per pulsar approximately
	  evenly distributed over $5$ years. Each pulsar has a white timing
	  noise of $100$ns. The true GW amplitude and slope are shown as a
	  ``+'' with an arrow, and the maximum likelihood values are shown as
	  ``x''.  The contours are in steps of $\sigma$, with the inner one at
	  $1\sigma$. The $1\sigma$ contour of the Gaussian approximation is also
	  shown.}
	  \label{fig:mcmc-white}
      \end{figure}

\begin{figure}
      \includegraphics[width=0.5\textwidth]{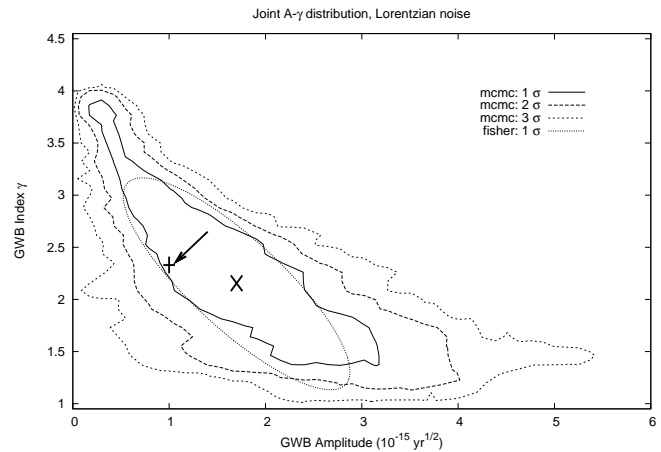}
      \caption{Same as in Fig.~\ref{fig:mcmc-white}, but the mock data is generated and
               analysed using Lorentzian timing noise. Overall timing noise
               amplitude and characteristic frequency $f_0$ are taken
               to be $100$ns and $1 \hbox{yr}^{-1}$ for each pulsar. }
	  \label{fig:mcmc-lorentzian}
      \end{figure}

\begin{figure}
      \includegraphics[width=0.5\textwidth]{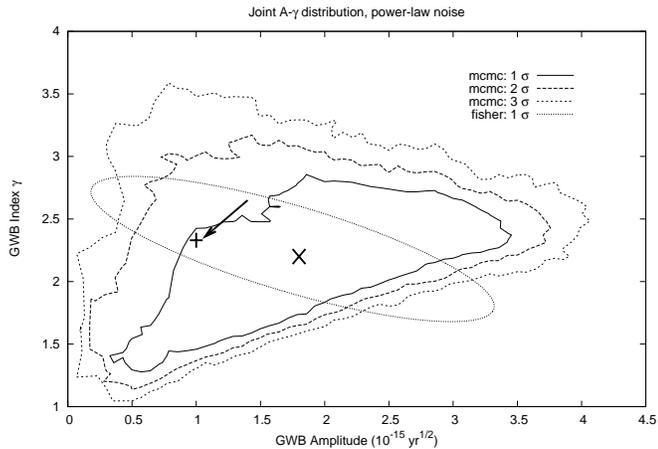}
      \caption{Same as in Fig.~\ref{fig:mcmc-white}, but the mock data is generated and
	       analysed using power-law timing noise. Overall timing noise
	       amplitude and spectral index $\gamma_i$ are taken to be $100$ns
	       and $1.5$ for each pulsar. For all $\gamma_i$, a prior
	       distribution according to Eq.~\ref{eq:prior}.}
	  \label{fig:mcmc-power}
      \end{figure}

      Our algorithm is tested on several datasets in the following way:\newline
      The mock datasets were generated with parameters resembling an experiment
      of $20$ pulsars, with observations approximately every $5$ weeks for $5$
      years. The pulsar timing noise was set to an optimistic level of $100$ ns
      each (rms timing residuals). In all cases the level of GWB has been set to
      $A=10^{-15} \hbox{yr}^{1/2}$, with $\gamma=7/3$. This level of GWB is an
      order of magnitude smaller than the most recent upper limits of this
      type\citep{Jenet-2006}. We then analyze this mock data using the MCMC
      method.  In Figs \ref{fig:mcmc-white}---\ref{fig:mcmc-power} we see
      examples of the joint $A$---$\gamma$ probability distribution, obtained by
      these analyses. For each dataset we also calculate the maximum likelihood
      value of all parameters using a conjugate directions search. The algorithm
      gives results consistent with the input parameters (i.e., they recover the
      amplitude and the slope of the GWB within measurement errors). This was
      observed in all our tests.

      For all datasets we also calculated the Fisher information matrix, a
      matrix consisting of second-order derivatives to all parameters, at the
      maximum likelihood points. We can use this matrix to approximate the
      posterior by a multi dimensional Gaussian, since for some particular models
      this approximation is quite good. The Fisher information matrix can be
      calculated in a fraction of the time needed to perform a full MCMC
      analysis. For all datasets we have plotted the $1\sigma$ contour of the
      multi dimensional Gaussian approximation.

      As an extra test, we have also used datasets generated by the popular
      pulsar timing package tempo2 \citep{Hobbs06} with a suitable GWB
      simulation plug-in (Hobbs et al., in preparation). We were able to
      generate datasets with exactly the same parameters as with our own
      algorithm, provided that the timing noise was white.  We have confirmed
      that those datasets yield similar results when analysed with our
      algorithm.

      An important point is that that the spectral form of the timing noise has
      a large impact on the detectability of the GWB. For a red Lorentzian
      pulsar timing noise there is far greater degeneracy between the spectral
      slope and amplitude in the timing residual data for the GWB than for white
      pulsar timing noise, and thus the overall signal-to-noise ratio is
      significantly reduced by the red component of the timing noise.
      
  \subsection{Multiple datasets, same input parameters} \label{sec:multipletests}
      To estimate the robustness of our algorithm, we also perform a maximum
      likelihood search on many datasets:\newline
      (a) We generate a multitude  of mock timing-residual data for the same PTA
      configurations as in Sec.~\ref{sec:singletests}, with white timing
      noise.\newline
      (b) For every one of these datasets we calculate the maximum likelihood
      parameters using the conjugate directions search. The ensemble of maximum
      likelihood estimators for ($A,\gamma$) should be close to the true values
      used to generate the timing-residuals.\newline
      
\begin{figure}
      \includegraphics[width=0.5\textwidth]{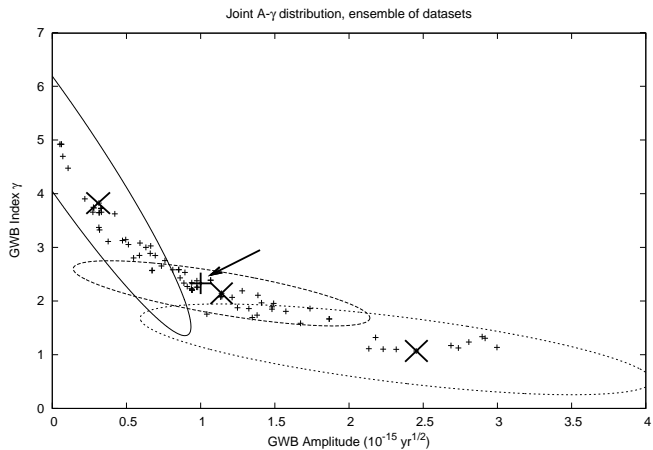}
      \caption{The maximum likelihood values for an ensemble of realisations of
	mock datasets, all with the same model parameters: $100$ ns white noise,
	$20$ pulsars, and $100$ data points per pulsar approximately evenly
	distributed over $5$ years. The contours are confidence contours based
	on Fisher information matrix approximations of the likelihood function.}
	\label{fig:ensemble-white}
      \end{figure}

      The results of maximum likelihood search on many datasets is demonstrated
      in Fig.~\ref{fig:ensemble-white}. The points are the maximum
      likelihood values for individual datasets. It can be seen that the
      points are distributed in a shape similar, but not identical, to
      Fig.~\ref{fig:mcmc-white}: some points are quite far off from the input
      parameters. In order to test the validity of the results, we calculate the
      Fisher information matrix at the maximum likelihood points, and show the
      $1\sigma$ contour of the multidimensional Gaussian approximation based on
      the Fisher information matrix for three points.  We wee that the error
      contours do not exclude the true values at high confidence, even though
      the Fisher matrix does not yield a perfect representation of the error
      contours (the true posterior is not perfectly Gaussian), and we have a
      posteriori selected outliers for 2 of the 3 cases.

    \subsection{Parameter studies} \label{sec:pstudies}
\begin{figure}
      \includegraphics[width=0.5\textwidth]{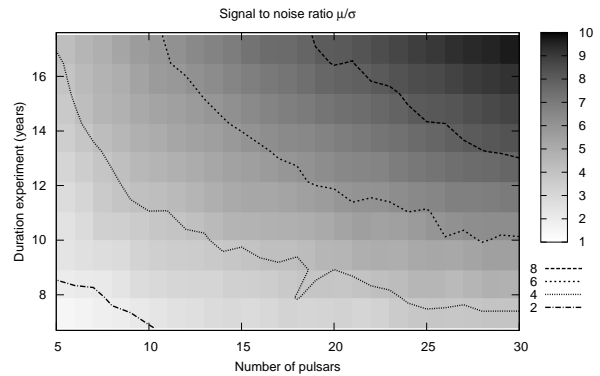}
      \caption{Density plot of the signal to noise ratio $\mu/\sigma$
	for different realisations of timing-residuals. We have assumed monthly
	observations of pulsars with white timing noise of $100$ ns each. The
	GWB amplitude has been set to $10^{-15} \hbox{yr}^{1/2}$.}
	\label{fig:ps-kT}
      \end{figure}

\begin{figure}
      \includegraphics[width=0.5\textwidth]{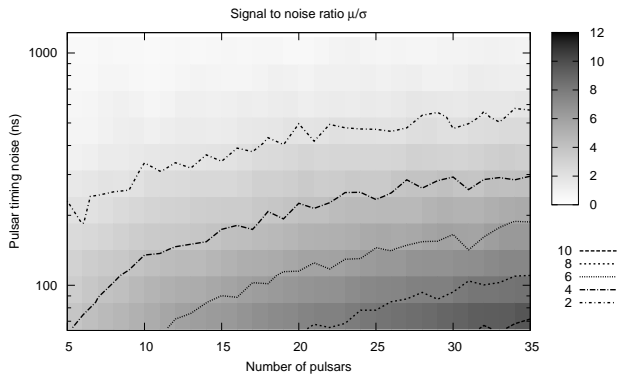}
      \caption{Density plot of the signal to noise ratio $\mu/\sigma$
	for different realisations of timing-residuals. We have assumed $100$
	data points per pulsars, approximately evenly distributed over a period
	of $7.5$ years. The GWB amplitude has been set to $10^{-15}
	\hbox{yr}^{1/2}$.}
	\label{fig:ps-kN}
      \end{figure}

\begin{figure}
      \includegraphics[width=0.5\textwidth]{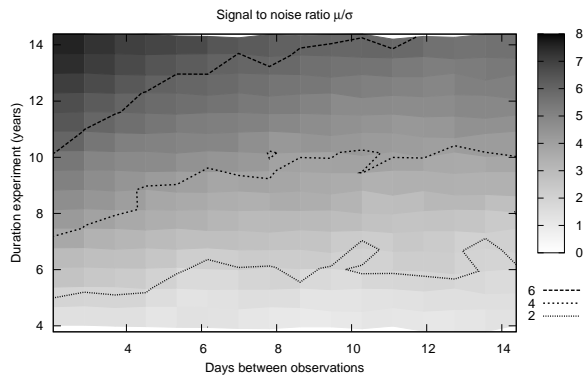}
      \caption{Density plot of the signal to noise ratio $\mu/\sigma$
	for different realisations of timing-residuals. We have used a constant
	GWB amplitude of $10^{-15} \hbox{yr}^{1/2}$ and $20$ pulsars.}
	\label{fig:ps-iT}
      \end{figure}

\begin{figure}
      \includegraphics[width=0.5\textwidth]{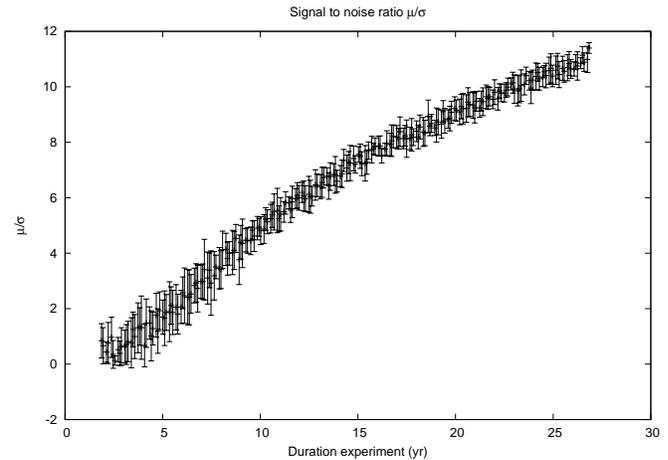}
      \caption{Density plot of the signal to noise ratio $\frac{\mu}{\sigma}$
	for different realisations of timing-residuals. We have used $20$
	pulsars with white pulsar timing noise levels of $100$ ns each, with
	monthly observations. The GWB amplitude has been set to $10^{-15}
	\hbox{yr}^{1/2}$.
	The points and error bars are the mean and standard deviation of $10$
	realisations.}
	\label{fig:ps-TT}
      \end{figure}

\begin{figure}
      \includegraphics[width=0.5\textwidth]{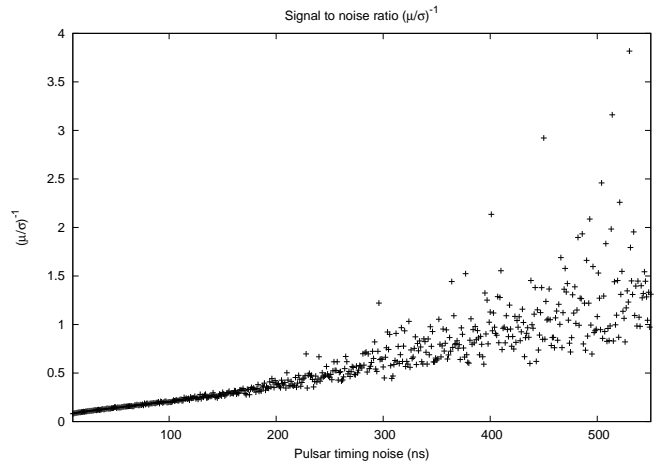}
      \caption{Plot of one over the signal to noise ratio
	$(\mu/\sigma)^{-1}$ with respect to the pulsar timing
	noise for an experiment of $5$ years, $20$ pulsars and monthly
	observations. The GWB amplitude has been set to $10^{-15}
	\hbox{yr}^{1/2}$.}
	\label{fig:ps-NN}
      \end{figure}

\begin{figure}
      \includegraphics[width=0.5\textwidth]{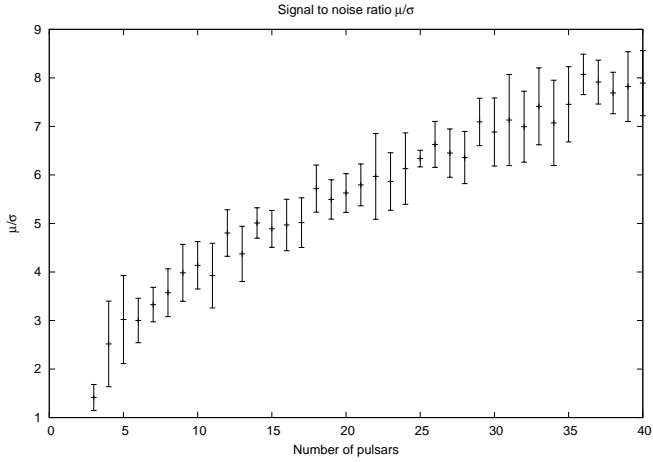}
      \caption{Plot of the signal to noise ratio $\frac{\mu}{\sigma}$ with
	respect to the number of observed pulsars. The white timing noise of
	each pulsar has been set to $100$ ns and the observations were taking
	every 2 months for a period of $7.5$ years. The GWB amplitude has been
	set to $10^{-15} \hbox{yr}^{1/2}$.
	The points and error bars are the mean and standard deviation of $10$
	realisations.}
	\label{fig:ps-kk}
      \end{figure}

\begin{figure}
      \includegraphics[width=0.5\textwidth]{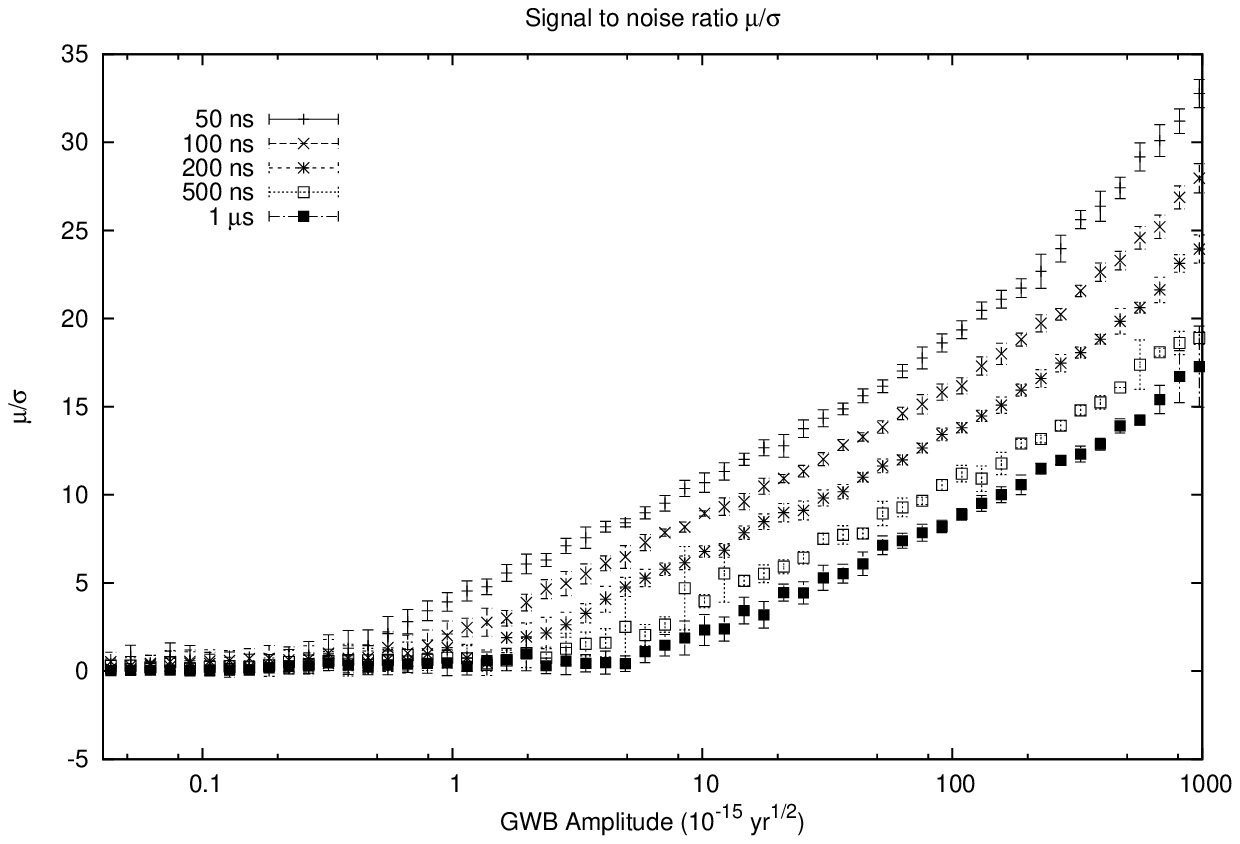}
      \caption{Several plots of the signal to noise ratio $\frac{\mu}{\sigma}$
	with respect to the level of the GWB amplitude. The number of pulsars
	was set at $20$, with bi-weekly observations for a period of $5$ years.
	The pulsar noise levels were set at $50$, $100$, $200$, $500$, $1000$ ns
	for the different plots.
	The points and error bars are the mean and standard deviation of $10$
	realisations.}
	\label{fig:ps-AA}
      \end{figure}

      To test the accuracy of the algorithm, and to provide suggestions for
      optimal PTA configurations, we conduct some parameter studies on
      simplified sets of mock timing-residuals:\newline
      (a) We generate many sets of mock timing-residuals for the simplified case
      of white pulsar timing noise spectra, all with the same white noise
      amplitude. The datasets are timing-residuals for some number of  millisecond pulsars
      which are positioned randomly in the sky. We parametrize the GWB by Eq.\
      (\ref{eq:sd2}). We then generate many sets of timing-residuals, varying
      several parameters [i.e., timing noise amplitude (assumed the same for all pulsars),
      duration of the experiment, and number of pulsars].\newline
      (b) For each of the mock datasets we approximate the likelihood function
      by a Gaussian in the GWB amplitude $A$, with all other parameters fixed to
      their real value. We use $A$ as a free parameter since it represents the
      strength of the GWB, and therefore the accuracy of $A$ is a measure of the
      detectability. All other parameters are fixed to keep the computational
      time low, but this does result in a higher signal to noise ratio than is
      obtainable with a full MCMC analysis.\newline
      (c) For this Gaussian approximation, we calculate the ratio
      $\mu\over\sigma$ as an estimate for the signal to noise ratio, where $\mu$
      is the value of $A$ at which the likelihood function maximizes, and
      $\sigma$ is the value of the standard deviation of the Gaussian
      approximation. Our results, represented as signal-to-noise contour plots
      for pairs of the input parameters, can be seen in
      Fig.~\ref{fig:ps-kT}---Fig.~\ref{fig:ps-AA}.

  \subsection{Comparison to other work}
    More then a decade ago, \citet{McHugh} used a Bayesian technique to produce
    upper limits on the GWB using pulsar timing\footnote{We thank the anonymous
    referee for attracting our attention to this paper.}. We found the
    presentation of this work rather difficult to follow. Nonetheless, it is
    clear that the analysis presented here is more general than that of McHugh
    et al.: we treat the whole pulsar array, and not just a single pulsar; we
    take into account the extreme redness of the noise and develop the formalism
    to treat the systematic errors like quadratic spindown. 

    Simultaneously with our work, a paper by Anholm et al. (2008, A08) has
    appeared on the arxiv preprint service. Their approach was to construct a
    quadratic estimator (written explicitly in the frequency domain), which aims
    to be  optimally sensitive to the GWB.  This improves on the original
    non-quadratic estimator of J05. However, a number of issues important for
    the pulsar timing experiment remained unaddressed, the most important among
    them the extreme redness of the GWB and the need to subtract consistently
    the quadratic spindown. 
  \section{Conclusion} \label{sec:conclusions}
    In this paper we have introduced a practical Bayesian algorithm for
    measuring the GWB using Pulsar Timing Arrays. Several attractive features of
    the algorithm should make it useful to the PTA community:\newline
    (1) the ability to simultaneously measure the  amplitude and slope of
    GWB,\newline
    (2) its ability to deal with unevenly sampled datasets, and \newline
    (3) its ability to treat systematic contributions of known functional form.
    From the theoretical point of view, the algorithm is guaranteed to extract
    information optimally, provided that our parametrization of the timing noise
    is correct.

    Test runs of our algorithm have shown that the experiments signal-to-noise
    (S/N) ratio strongly decreases with the redness of the pulsar timing noise,
    and strongly increases with the duration of the PTA experiment. We have also
    charted the S/N dependence on the number of well-clocked pulsars and the
    level of their timing noise. These charts should be helpful in the design of
    the optimal strategy for future PTA observations.

  \section{Acknowledgements} \label{sec:acknowledgements}
    We thank Dan Stinebring, Dick Manchester, George Hobbs, Russell Edwards,
    Rick Jenet, Ben Stappers, Jason Hessels, and Matthew Bailes for insightful
    discussions about the precision pulsar timing. RvH and YL  thank ATNF for
    its annual hospitality. This research is supported by the Netherlands
    organisation for Scientific Research (NWO) through VIDI Grant $639.042.607$.

    \clearpage
  \section*{Appendix A}
    In this Appendix we show explicitly how to perform marginalization over
    the nuisance parameters $\vec{\xi}$ in Eq.~(\ref{deltareal2}), rewritten 
    here for convenience:
    \begin{eqnarray}
      P(\sigma, \vec{\xi}|\vec{\delta
      x})&=& {M\over \sqrt{\det{C}}}\exp \left[-{1\over 2}
      (\vec{\delta x}-F\vec{\xi}) C^{-1} (\vec{\delta x}-F\vec{\xi}) \right]
      \nonumber\\
      & &\times P_0(\sigma, \vec{\xi}),
      \label{eq:apppb1}
    \end{eqnarray}
     From here on we
    will assume that $P_0$ is independent of $\vec{\xi}$ (a flat prior). All
    values are therefore equally likely for all elements of $\vec{\xi}$ prior to
    the observations. This assumption is also implicitly made in the frequentist
    approach when fitting for these kinds of parameters as is done in popular
    pulsar timing packages. We now perform the marginalisation:
    \begin{equation}
      P(\sigma|\vec{\delta x}) = \int P(\sigma, \vec{\xi}|\vec{\delta x})\hbox{d}^m\xi,
      \label{eq:marginalise}
    \end{equation}
    where $m$ is the dimensionality of of $\vec{\xi}$. The idea now is to
    rewrite the the exponent $E$ of Eq.~(\ref{eq:apppb1}) in such a way that we
    can perform a Gaussian integral with respect to $\vec{\xi}$ (we have to get
    rid of the $F$ in front of $\vec{\xi}$). Therefore, we will expand $E$ and
    complete the square with respect to $\xi$:
    \begin{eqnarray}
      E &=& \left( \vec{\delta x}-F\vec{\xi} \right)^{T}C^{-1}\left( \vec{\delta
      x}-F\vec{\xi} \right) \nonumber\\
      &=& \vec{\delta x}^{T} C^{-1} \vec{\delta x}
	  - 2 \vec{\xi}^{T}F^{T} C^{-1} \vec{\delta x}
	  + \vec{\xi}^{T}F^{T} C^{-1} F \vec{\xi} \nonumber\\
      &=& \vec{\delta x}^{T} C^{-1} \vec{\delta x}
	  +\left( \vec{\xi}-\vec{\chi} \right)^{T} F^{T}C^{-1}F
	   \left( \vec{\xi}-\vec{\chi} \right) \nonumber\\
      & & -\vec{\chi}^{T}F^{T}C^{-1}F\vec{\chi},
    \end{eqnarray}
    where we have used the substitution:
    \begin{equation}
      \vec{\chi} = \left( F^{T}C^{-1}F \right)^{-1}F^{T}C^{-1}\vec{\delta x}.
    \end{equation}
    Using this, we can write the $\vec{\xi}$ dependent part of the integral of
    Eq.~(\ref{eq:marginalise}) as a multi dimensional Gaussian integral:
    \begin{eqnarray}
      I &=&\int\exp\left(\frac{-1}{2} \left( \vec{\xi}-\vec{\chi} \right)^{T} F^{T}C^{-1}F
	     \left( \vec{\xi}-\vec{\chi} \right) \right)\hbox{d}^m\xi \nonumber\\
	&=& \left( 2\pi \right)^{m}\det\left( F^{T}C^{-1}F \right)^{-1}.
    \end{eqnarray}
    From this it follows that:
    \begin{eqnarray}
      P(\sigma|\vec{\delta x})&=&{M'\over \sqrt{\det(C)
	    \det(F^{\rm T} C^{-1} F)}}\\ & &\times \exp\left[-{1\over
	    2}\vec{\delta x}\cdot C^{\prime} \vec{\delta x}\right],\nonumber
      \label{premoved2}
    \end{eqnarray}
    where we have absorbed all constant terms in the normalisation constant
    $M'$, and where we have used:
    \begin{equation}
      C^{\prime}=C^{-1}-C^{-1}F(F^{\rm T} C^{-1}F)^{-1}F^{\rm T}C^{-1}.
    \end{equation}

%-- main text ------------------------------------------------------------------

%-- BIBLIOGRAPHY ----------------------------------------------------------------
  \bibliographystyle{mn2e.bst}
  \bibliography{ptaarticle}
%-- bibliography ----------------------------------------------------------------

\end{document}